# Detecting Bots Based on Keylogging Activities

Yousof Al-Hammadi and Uwe Aickelin Department of
Computer Science and Information Technology, The University
of Nottingham,
Nottingham, UK NG8 1BB
Email: *{yxa,uxa}*@cs.nott.ac.uk

*Abstract*—A bot is a piece of software that is usually installed on an infected machine without the user's knowledge. A bot is controlled remotely by the attacker under a Command and Control structure. Recent statistics show that bots represent one of the fastest growing threats to our network by performing malicious activities such as email spamming or keylogging. However, few bot detection techniques have been developed to date. In this paper, we investigate a behavioural algorithm to detect a single bot that uses keylogging activity. Our approach involves the use of function calls analysis for the detection of the bot with a keylogging component. Correlation of the frequency of function calls made by the bot with other system signals during a specified time-window is performed to enhance the detection scheme. We perform a range of experiments with the spybot. Our results show that there is a high correlation between some function calls executed by this bot which indicates abnormal activity in our system.

## I. INTRODUCTION

For some time now, computers face different types of attacks by malicious programs such as viruses and worms[11][22]. A more recent threat is the presence of large numbers of compromised machines, known as *bots*, working in a coordinated manner [13]. A *bot*, a term derived from robot, is a piece of malicious software that is installed on a user machine, usually without his knowledge. This malicious software is programmed to respond to various instructions remotely by the attacker through Command and Control (*C&C*) structure, often using the Internet Relay Chat (*IRC*) network as a communication channel. These instructions command the bot on the infected machine to perform malicious activities. The malicious activities include vulnerability scans to spread the bot to other systems, email spamming, keylogging, packet sniffing, phishing, rootkits and identity theft.

A bot spreads and propagates to other hosts by exploiting known vulnerabilities in operating systems and applications. The target host is infected by different means ranging from worms, email viruses or phishing [21]. Once the bot is installed on a victim's machine, it changes the system configuration to start itself each time the system boots. A bot might have the functionality to spread itself by sending out more emails or scanning more computers, thus, infecting other vulnerable machines. After that, the bot connects to the IRC server and joins the specified Command and Control channel. Having joined the channel, the bot either executes the channel topic as a default command or remains inactive waiting for botmaster commands. The botmaster communicates with the bot through IRC protocol[1]. The IRC protocol is a preferred due to its flexibility in the management and control of bots. Additionally, the IRC protocol provides the attackers with anonymous control over their bots. The botmaster can also control the bots using different types of communications such as the HTTP protocol or through Peer to Peer networks.

Initially, bots were used to coordinate attacks across a network of bot-infected machines. Nowadays, most bots are implemented with keylogging features. Keylogging is a mean of intercepting and subversively monitoring the user activities such as typing keystrokes and mouse clicking. The intercepted keystrokes are either saved to a log file or sent directly to the botmaster. The log file can be sent to the attacker through email, ftp or accessed remotely by the attacker. Other new features added to the keylogging bot are the ability to capture screen shots, and mouse logging [14].

Keylogging represents a serious threat to the privacy and security of our systems. This is because the keylogger program can collect the user's personal information, passwords, credit cards or other sensitive information. Unlike other attacks performed by the bot, keylogging is difficult to detect as it runs in hidden mode. Many Anti-Virus packages cannot detect a stealthy keylogger running on the system. The user has no way to determine if his machine is running a keylogger, therefore, he could easily become a victim of the identity theft.

The focus of bot detection research is the analysis of network traffic. To the best of our knowledge, no attempt has been made within bot detection research to detect a single bot by monitoring Application Programming Interface (*API*) function calls. In this paper, we present an algorithm to detect a single bot in the system based on correlating different behaviours by monitoring specified API function calls executed by the bot to perform keylogging activity. Invoking these functions withing specified time window might represent a security risk to computer systems. For example, calling *GetKeyboardState* or *GetAsyncKeyState* by a program and writing data to a file using the *WriteFile* function call usually indicates a keylogging activity. In addition, the bot is designed to send the intercepted keystrokes to the attacker, therefore, we may notice a large volume of outgoing traffic during this period. Correlation of the frequency of function calls generated by the bot during a specified time-window could indicate abnormal activity in our system. Overall, we believe that tracking and correlating the keyboard events with other behaviour data such as accessing files or sending packets will enhance the process of bot detection.

The aim of this paper is to investigate the effect of correlating different behaviours of a single bot represented by API function calls within specified time-window. We focus on three types of bot behaviour: keylogging activity, file access and outgoing traffic. Our results show that correlating different behaviours of a single bot enhance the bot detection mechanism.

The remainder of this paper is structured as follows: section two presents different algorithms used to detect bots and keylogging activities. It also shows the problems which face the current detection systems. Section three discusses the design and implementation of our bot detection. In addition, it explains different types of experiments that we perform to our bot detection. We discuss our results in section four. We conclude and present our future work in section five.

## II. RELATED WORK

There are only a few existing techniques for bots detection. Most of these techniques use signature-based detection by analysing network traffic [10][9]. Although analysing network traffic using a signature-based approach is a useful mechanism for bot detection, it becomes more difficult if the botmaster's commands are encrypted. Different research performed by Binkley [7] uses anomaly-based detection to detect the behaviour of the bot. The anomaly detection technique looks for the deviation from a defined normal traffic. In this section, we will present current related work in bot detection techniques and the associated problems.

Recent work by Barford [5] represents a good introduction to understanding and analysing the behaviours of the bots. Most of the research conducted in this area concentrates on detecting botnets rather than an individual bot [10][9][2] and, to the best of our knowledge, little research has been performed in this area. Freiling et al. [10][13] use a non-productive resource such as honeypot to collect bot binaries. Their approach is based on allowing the infected honeypot to emulate the bot activities and analysing network traffic to shut down the remote control network. Although honeypots allow administrators to look at security events in more detail, they cannot detect these events without receiving activity directed against them [3]. In addition, the process of emulating the bot action to penetrate the remote network can be discovered if the botnet size is relatively small. To avoid these problems, our work focuses on monitoring *API* function calls generated by the bot and correlates these function calls within a specified time-windows to detect malicious activities. In addition, host-based detection is used to investigate the presence of the bot in our system.

Cooke et al. [9] detect bots by analysing the communications between bots, and the communications between bots and their controllers. They also investigate the bots payloads using the pattern matching of known bot commands and look for the behavioural characteristics of bots that differ from non-human characteristics. They conclude that bots can run on non-standard ports and that analysing encoded packets is very costly on high throughput networks. Hence, there are no simple characteristics of the bots communication channels that can be used for detection. They also discuss the approach of detecting bots by their propagation or attack behaviour by correlating data from different sources. However, they have not designed a correlation algorithm to show results of this investigation. In this paper, we present one approach of detecting bots based on correlating the frequency of different bot's behaviour such as keyboard events, files access and the amount of outgoing traffic.

An approach for the analysis of IRC usage by bots is presented by Stephane Racine [21]. This approach detects bots by finding inactive clients through monitoring IRC *PONG* messages and assigning them to a connection. The active clients are then classified according to the channel that they join. This approach is successful in detecting idle IRC activity, but suffers from high false positive rates. In addition, searching for IRC patterns can be costly when inspecting every packet and could slow the detection mechanism. Furthermore, applying pattern matching is difficult when data is encrypted [9]. We believe that monitoring and correlating different *API* function calls will enhance the process of bots detection through this correlation.

Research in keylogging has shown that is difficult to detect but the literature on this topic is also sparse. TAN [23] suggests disabling some function calls used by the keylogger such as *SetWindowsHookEx*, GetKeyboardState and *GetKeyState*. This can prevent the proper functioning of keyloggers. However, disabling these functions will prevent legitimate software from using these functions. In our work, we monitor calls to some of these functions and other functions. Monitoring calls to these functions will not affect their use by legitimate programs.

Other research suggests that embedding a sequence of random characters between successive keys typed on the browser will make the keylogging process difficult [12]. Another group [6] disassembles all running processes searching for *SetWindowsHookEx* used by some keyloggers. One problem with this method is that the keylogger developers can use different methods to log the user activities other than using *SetWindowsHookEx*. In addition, disassembling all processes searching for *SetWindowsHookEx* is a tedious task. Our approach is based on monitoring selected *API* function calls for all running processes in our system in user mode. By monitoring these functions, we will avoid the problems mentioned above.

## III. BOT DETECTION

### A. Introduction

Existing research techniques detect the presence of bots via network monitoring. Rather than attempting to detect bots via network monitoring, our work focuses on a single bot detection on a machine by monitoring and correlating different activities on the system represented by executing different API function calls that may indicate the presence of a bot on the system.

The API function calls executed by the bot is monitored in a *user-based* environment. Our monitoring program intercepts API function calls in the *user-based* environment. Direct invocation kernel-based API functions will not be monitored by our intercepting program. Monitoring kernel-based API function calls will be investigated in our future work.

In order to detect the bot in the system different bot behaviours are correlated to have a high correlation value, which is represented by Spearman's Rank Correlation (*SRC*) value [8]. In our case, if the Spearman's Rank Correlation value exceeds the threshold level of 0.5, we have a high correlation between the two different behaviours which may reflect malicious activity in our system. The threshold level of 0.5 or more represents a strong correlation between two events according to the Speaman's Rank Correlation algorithm. We hypothesize that one behaviour may not be enough to detect malicious activity. This is explored in section IV. For example, one behaviour of the bot is to send the intercepted information from the keylogging process to the botmaster once the botmaster issues the keylogging command. The intercepted information is sent to the botmaster if the user of the infected machine hits [ENTER] key, or closes the active window. This action represents normal behaviour. Correlating different actions enhance the process of detection. We are aware that different keyloggers use different techniques to intercept and store the keystrokes. Our work examines a keylogging activity as our goal is to detect a bot rather than a keylogger.

### B. Aims

The aim of our experiments is to verify the notion that correlating different behaviours of a single process which produces multiple API function calls within a specified time-window, indicates abnormal activity. In addition, we apply the monitoring and correlation scheme to a normal application (e.g. mIRC client) to verify that the normal application executes different function calls from the malicious process which results in having different correlation value.

### C. Design and Implementation

In our research, we focus on monitoring selected API function calls executed by bots that perform the keylogging task and send the intercepted keystrokes directly to the IRC channel. To accomplish this task, we implement a program to monitor some API function calls executed by the bot when receiving commands from the botmaster. We focus on three types of API function calls:

- **Communications Functions (CommFunc)**: *socket*, *send*, *recv*, *sendto*, *recvfrom*, and *IcmpSendEcho* [20].
- **File Access Functions (FileAccess)**: *CreateFile*, *OpenFile*, *ReadFile*, and *WriteFile*[18].
- **Keyboard State Functions (KeyboardState)**: *GetKeyboardState*, *GetAsyncKeyState*, *GetKeyNameText*, and *keybd_event*[19].

We have implemented a 'hook' program to capture the API functions executed by the bot. Hooking API functions is the process of intercepting events (messages, keystrokes, mouse) before they reach an application[15][16].

In our work, we captured selected API functions such as GetKeyboardState, and GetAsyncKeyState used by bots which implement keylogging feature. For example, the spybot[5] is used for its ability to intercept the user's keystrokes by invoking GetAsyncKeyState. We search for all the running processes in our system and inject our hooking program into the running processes. An API hook is based on modifying the process Import Address Table [4] to point to the replacement function instead of the original function. Thus, we were able to capture the functions made by the bot when it receives commands from the botmaster.

We store the captured functions in a log file for further processing. We use a Spearman's Rank Correlation formula [8] to find the correlation between different behaviours of the bot such as intercepting the user keystrokes and sending it directly to the IRC channel within specified time-window. In addition, we also correlate the events of intercepting the user keystrokes and file access. Our results show that the combination of these correlated events can indicate suspicious activity in our system.

The algorithm of detecting the bot is described in Algorithm 1.

**Algorithm 1**: Bot Detection Algorithm using Spearman's Rank Correlation (SRC)

**if** *KeyboardState function(s) is executed (i.e. keylogging activity)* **then**
    **if** *SRC[KeyboardState,CommFunc] > Threshold and SRC[KeyboardState,FileAccess] > Threshold* **then**
        Strong detection
    **else if** *SRC[KeyboardState,CommFunc] < SRC[KeyboardState,FileAccess] < Threshold* **then**
        Weak detection
    **else if** *(SRC[KeyboardState,CommFunc] < Threshold and SRC[KeyboardState,FileAccess] > Threshold) or (SRC[KeyboardState,CommFunc] > Threshold and SRC[KeyboardState,FileAccess] < Threshold)* **then**
        Normal detection
**else**
    No detection and normal activity is considered
**end**

### D. Architecture

To perform our experiments, we set up a small virtual IRC network on a VMWare machine. The VMWare machine runs under a Windows XP P4 SP2 with a 2.4GHz processor and 1GB RAM. The virtual IRC network consists of two machines. One machine runs Windows XP Pro SP2 and it is used as an IRC server. The other machine runs Windows XP Pro SP2 as an infected machine with spybot [5]. We do not have to have a large network to implement our algorithm as our work based on detecting the behaviour of a single bot on a machine.

### E. Experiments

We have performed five experiments to verify our notion. In the first experiment (*E1*), we allow the spybot to connect to the IRC server and join the channel without receiving any commands from the botmaster. In the second experiment (*E2*), we follow the same procedure as in the first experiment, but in this case the botmaster issues different commands to the bot, excluding the keylogging activity. Note that our target machine in these experiments is an idle infected machine. That is, the user does not use the infected machine for any activity.

In the third experiment, we allow the bot to connect to IRC server and join the specified channel. The bot on the infected machine monitors the user's typing activity, but does not send any information to the botmaster. We monitor two scenarios of typing. In the first scenario (*E3.1*), the user types long sentences while in the second scenario (*E3.2*), the user types short sentences. By monitoring two typing scenarios, we are able to show the effect of different user's activity on our detection scheme.

In the fourth experiment, once the bot connects to the IRC server and joins the channel, the botmaster starts the keylogging activity. The same procedure is taken as in the third experiment where we have two scenarios of typing: long sentences (*E4.1*) and short sentences (*E4.2*).

The fifth and the final experiment (*E5*) involves applying the monitoring program to another application (*mIRC* client [17]) to verify that *mIRC* client behaves differently from the bot.

Each experiment is performed five times which is sufficient as the results from the repeated experiments produce only small variations by using Chebyshevs Inequality due to network delay and through using VMWare. Therefore, we select a random experiment from the repeated experiments as the base experiment. Each experiment runs for 15 minutes in order to collect a reasonable number of function calls which reflect most of botmaster execution commands. The monitored API functions are saved into a log file. After that, we use a Spearman's Rank Correlation (SRC) method to correlate different behaviour of the bot based on the frequency of API function calls executed by the bot in our system within a specified time-window. In our experiments, a time-window of ten seconds is used between function calls samples. We notice that monitoring function calls for a time-window of 60 seconds will have variant idle periods depends on the bot activity. An idle period is where no bot activity is detected and zero values are assigned. Therefore, using a time-window of ten seconds reduces the idle periods suitably.

Our assumption is that calling GetAsyncKeyState or GetKeyboardState functions by an unknown running program may represent abnormal keylogging activity in our system. However, we consider that calling these functions generate only a 'weak' alert because other programs may use the same API calls. Therefore, we use Spearman's Rank Correlation to correlate different types of bot behaviour which enhances our detection algorithm to form a 'strong' alert.

The Spearman's Rank Correlation correlates two different data sets. The first data set is the outgoing traffic from our system (i.e., total number of bytes sent to the botmaster every ten seconds) and the frequency of GetAsyncKeyState function calls generated. The second data set is the frequency of GetAsyncKeyState function calls and the frequency of WriteFile function calls generated. These function calls are important for monitoring bot behaviour because their invocation represents

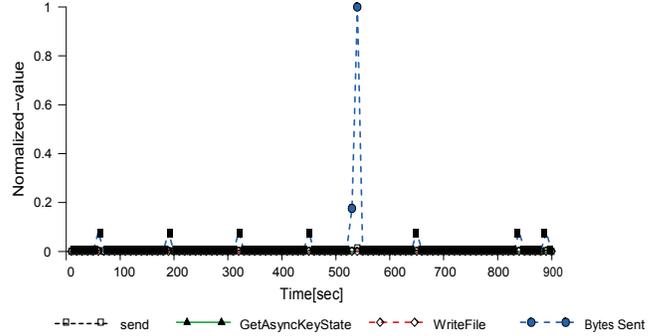

Fig. 1. The results of experiment E1. The bot connects to the IRC server, joins the specified channel and remains inactive waiting for the botmaster's commands.

abnormal behaviour within our system.

## IV. RESULTS AND ANALYSIS

In this section, we analyse the results of the experiments described in Section III-E. For all experiments, the x-axis represents time in seconds while the y-axis represents the normalized value of functions. The normalized function frequency call values represent the total value we get during 10 seconds divided by the maximum value of the whole period (900 seconds). In addition, we use a line graph which connects the points to make our figures more readable.

In experiment *E1*, the bot is idle for the majority of the duration. This means that no API function calls are executed except the communication functions, specifically, *send* and *recv*, as shown in Figure 1. From Figure 1, we notice that it is difficult to detect the bot's behaviour as there is no activity in the system except the communications. We also notice that there is a burst in the outgoing traffic. This burst is generated due to spybot program which sends a bulk of words every specified time intervals.

In experiment *E2*, the botmaster issues commands such as *info*, *list* and *passwords* and the bot on the infected machine responds to these commands. Each time the botmaster issues a command, different API function calls are executed by the bot. In this experiment, we noticed an increased amount of outgoing traffic compared to experiment *E1*. In addition, few WriteFile and ReadFile functions are generated during this experiment. Conversely, no GetAsyncKeyState function calls are generated, as shown in Figure 2.

The third experiment has two typing scenarios: (1) Long sentences (*E3.1*) and (2) Short sentences (*E3.2*). Figure 3 represents the long sentences scenario *E3.1*. We notice that even though we have many GetAsyncKeyState function calls executed by the bot, which indicates keylogging activity, there is almost no correlation between GetAsyncKeyState and WriteFile. This is because the WriteFile function call is rarely generated as it is only triggered when the user types long sentences. To save the long sentences, the user has to press the Enter key or close the application. In addition, no data is sent to the botmaster which reduces the correlation value between GetAsyncKeyState and the outgoing traffic. In scenario *E3.2*, the user of the infected machine types short sentences. We can see from Figure 4 that there is a high correlation between GetAsyncKeyState and WriteFile function calls. This situation is expected as each time the user types short sentences, the functions GetAsyncKeyState and WriteFile are called to intercept the user keystrokes and store them in a file. However, there is still no traffic sent out and hence there is no correlation with outgoing traffic.

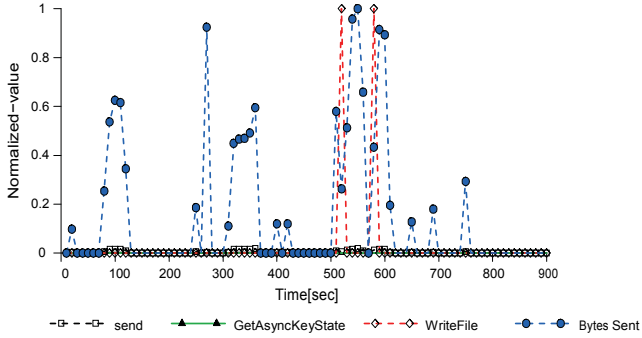

Fig. 2. The results of experiment E2. The bot receives commands from the botmaster. The amount of outgoing traffic increases as the bot responds to the botmaster's commands.

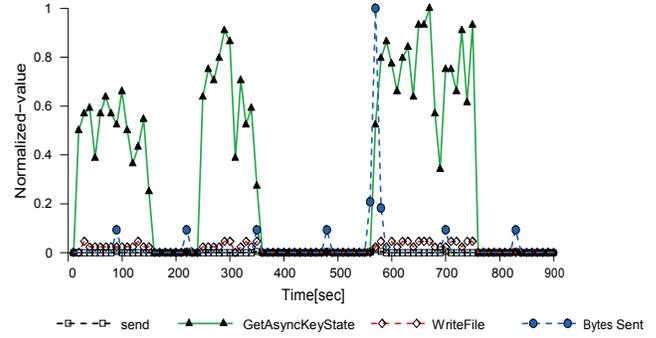

Fig. 4. The results from the third experiment - scenario E3.2. The botmaster has not activated the keylogger command. The user on the infected machine types short sentences.

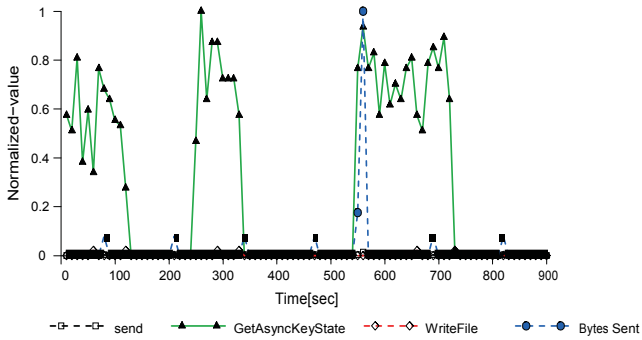

Fig. 3. The results from the third experiment - scenario E3.1. The botmaster has not activated the keylogger command. The user on the infected machine types long sentences.

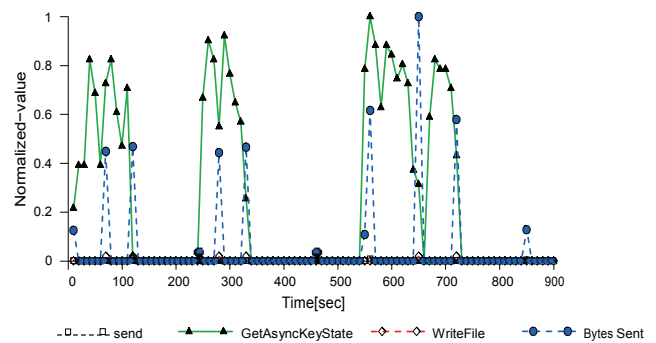

Fig. 5. The first scenario E4.1 in experiment four. The botmaster activates the keylogger. The user on the infected machine types long sentences.

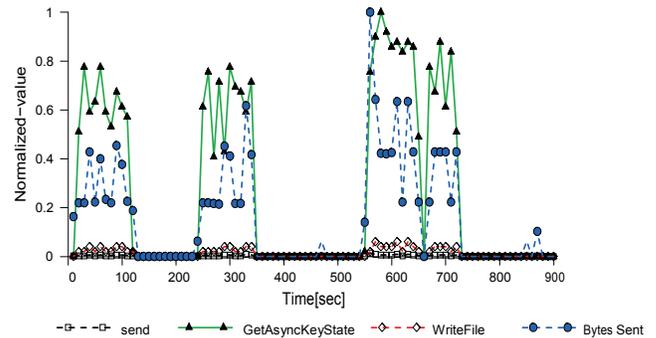

Fig. 6. The second scenario E4.2 in experiment four. The botmaster activates the keylogger. The user on the infected machine types short sentences.

In experiment 4, the botmaster starts the keylogging activity and the intercepted keystrokes are sent to the botmaster. In this case, we also have two typing scenarios: (1) Long sentences (*E4.1*) and (2) Short sentences (*E4.2*). In scenario *E4.1*, we expect there to be a high correlation between the outgoing traffic and GetAsyncKeyState. However, the result from Figure 5 shows that there is a low correlation between the two. This is because we correlate the two events (typing and saving to a file) in two different 10 second time intervals. In addition, the long sentences increase the idle time, and therefore reduce the correlation value. Moreover, a low correlation between GetAsyncKeyState and WriteFile is noticed. This situation is expected as the user types long sentences which call few WriteFile functions.

In the second scenario *E4.2*, the user types short sentences resulting in a high correlation between the outgoing traffic with the GetAsyncKeyState function and between the GetAsyncKeyState function and the WriteFile function as shown in Figure 6. The high correlation in both cases increases the amount of evidence for a bot spying on our system.

In addition, we test our monitoring program with the mIRC

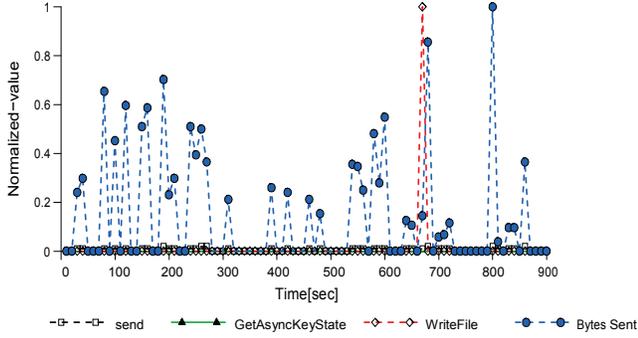

Fig. 7. The results from Experiment E5. The mIRC client connects to the IRC server. The client has normal conversation and simple commands with another client.

program. The result in Figure 7 is optimistic as the program did not call any GetAsyncKeyState or GetKeyboardState functions.

Table I represents the value of Spearman's Rank Correlation between the two data sets, (`GetAsyncKeyState`, `Bytes Sent`) and (`GetAsyncKeyState`, `WriteFile`), in each experiment. In this table, we have two sets of results. In the first set *S1*, we correlate all the captured data from our algorithm including the idle period. In this period, no activity is seen, therefore, we assign a zero value to this period. This is represented by the *with zero* column in Table I. In the second set *S2*, we remove all the idle periods which have zeros and apply the Spearman's Rank Correlation to the new data. The reason for having the two sets is that we notice that having the idle periods in our data increases the correlation value. This is because there are many places where no activity is noticed in both data sets, which may produce inaccurate correlation. Therefore, we wanted to investigate the effect of having no idle periods. Although we notice a reduction of the correlation value by 0.35 in most cases when we remove the idle periods, it gives us more accurate results.

The *API Keylogging Activity* column represents the situation where the process calls any function used to intercept the keystrokes such as GetAsyncKeyState, GetKeyboardState, GetKeyNameText and keybd_event. Calling these functions may indicate a keylogger activity. As a result, we classify our detection scheme into four cases:

- No detection (N/A): the case where no keylogging activity is detected.
- Weak detection (Weak): the case where a keylogging activity is detected but a low correlation is noticed in both data sets.
- Normal detection (Normal): the case where a keylogging activity is detected but a high correlation is noticed in one data set.
- Strong detection (Strong): the case where a keylogging activity is detected but a high correlation is noticed in both data sets.

As mentioned in section III-A, a high correlation is considered if the Spreaman's Rank Correlation value exceeds the threshold (0.5). Conversely, a low correlation value is considered if the Spearman's Rank Correlation value is below the threshold.

From Table I, we see a perfect correlation of GetAsyncKeyState and WriteFile function calls in experiment *E1*. The bot called neither of these functions during its inactive period. We also notice that there is a high Spearman's Rank Correlation value between the outgoing traffic (Bytes Sent) and GetAsyncKeyState because the amount of outgoing traffic is equal each time. This traffic belongs to the PONG message generated by the bot to avoid disconnection from the IRC server. Therefore, the correlation value is expected to be high as well. In experiment *E2*, the high Spearman's Rank Correlation value is due to the correlation of GetAsyncKeyState and WriteFile which are not invoked and zero values are assigned.

In experiment *E3.1*, we notice a call to GetAsyncKeyState which indicates abnormal activity. On the other hand, a low Spearman's Rank Correlation value is generated in both data sets. This situation is expected because the user types long sentences which make only a few calls to WriteFile. In addition no information is sent to the botmaster. As a result, a weak detection is indicated. Experiment *E3.2* detects a keylogging activity and generates a high correlation between GetAsyncKeyState and WriteFile executed by the bot due to typing short sentences. On the other hand, no information is sent to the botmaster which results in normal detection according to our classification.

Experiment *E4.1* shows similar activity to experiment *E3.1* where the user types long sentences, but the information is sent to the botmaster. We expect to have a high correlation between the outgoing traffic and GetAsyncKeyState function. The result shows there is no significant difference from experiment *E3.1*. This is because the bot sends the information when the user finishes typing long sentences. Experiment *E4.2* is the best case for detecting keylogging activity in our system. In this experiment, we detect the keylogging activity and we have high correlation values for both data sets which indicates abnormal activity running in our network.

The last experiment *E5* in Table I shows the result of the Spearman's Rank Correlation correlation on monitoring the mIRC client. Even though we have a high correlation value before and after removing idle periods on both experiments, we did not detect the use of keylogging function calls. The high correlation value between *outgoing traffic* and GetAsyncKeyState relates to the number of of idle periods due to the delay in responding to another client's messages.

In summary, we notice that some experiments produce high correlation values. There are many reasons for this. The first reason is that different events occur in different time-windows. Therefore, our algorithm produces inaccurate results. The second reason is that we have many idle periods in our data sets. The idle periods increase the correlation value which affect our detection scheme. In order to improve our detection scheme, we need to apply a more intelligent correlation scheme.

TABLE I
SPEARMAN'S RANK CORRELATION (SRC) VALUE WHICH REPRESENTS
THE CORRELATION BETWEEN TWO DATA SETS.

| Experiments | SRC(GetAsyncKey, Bytes Sent) | | SRC(GetAsyncKey, WriteFile) | | Keylog. Activity existence | API Detection confidence |
|---|---|---|---|---|---|---|
| | with zeros (S1) | without zeros (S2) | with zeros (S1) | without zeros (S2) | | |
| E1 | 0.863 | 0.671 | 1.000 | 1.000 | No | N/A |
| E2 | 0.648 | 0.498 | 0.967 | 0.897 | No | N/A |
| E3.1 | 0.509 | 0.183 | 0.559 | 0.172 | Yes | Weak |
| E3.2 | 0.423 | -0.003 | 0.928 | 0.618 | Yes | Normal |
| E4.1 | 0.506 | 0.189 | 0.560 | 0.089 | Yes | Weak |
| E4.2 | 0.927 | 0.579 | 0.957 | 0.663 | Yes | Strong |
| E5 | 0.594 | 0.499 | 0.983 | 0.958 | No | N/A |

## V. LIMITATION OF THE ALGORITHM

As our algorithm is based on bots detection by correlating different behaviours within specified time window, it is possible for the botmaster to evade this detection technique by allowing the bot to wait for a random time before performing another task. In this situation, we need to increase the time window and search for correlation of events within this time window. Increasing the time window may have a negative impact on our bots detection period.

Another important issue is that we focus on detecting a bot based on its keylogging activity. Combining other bot activities such as SYN attack or UDP attack with the keylogging activities can increase the correlation between different events.

## VI. CONCLUSIONS AND FUTURE RESEARCH

In this paper, we develop a program to monitor some API function calls of the spybot. We consider the execution of these functions within specified time-window as a security risk to our system. We highlight that looking at API function calls alone is not sufficient as other normal applications can call the same functions. As a result, the need for correlating different behaviour data is recommended. We use a Spearman's Rank Correlation method to correlate our captured data. Although the Spearman's Rank Correlation is a simple method to examine the correlation level, the results were promising. In addition, our results show that including idle periods in our correlation algorithm produces inaccurate results. This is due to the fact that most of the situations examined had a large number of idle periods which increases the correlation value. A more intelligent way of correlating the data is to remove the idle periods. Although removing the idle periods reduces the correlation value, it produces more acceptable results. It would be even better to remove only certain idle periods, something we will consider in our future research.

Other experiments show a weak detection decision (low correlation values). This is because different activities occur in different time-windows. For example, the typing process has a different time window than writing to a file or sending information to the attacker. We believe that choosing a correct time-window as well as the window size to correlate our data will have a large effect on our detection algorithm. Currently, we are using the Artificial Immune System approach for correlating different activities within the same time window. For future work, we will use the Artificial Immune system correlation algorithm to detect the Peer-to-Peer bots.

## ACKNOWLEDGMENT

The authors would like to thank Etisalat College of Engineering and Emirates Telecommunication Corporation - ETISALAT, for providing financial support for this work.